\documentclass[preprint,12pt]{elsarticle}

\usepackage{amssymb}
\usepackage{amsmath}

\usepackage{float}
\usepackage[hyphens]{url}
\usepackage{breakurl}
\usepackage{multirow}
\usepackage{makecell}
\usepackage{array}
\usepackage{subcaption}

\journal{Computers \& Graphics}

\begin{document}
\begin{sloppypar}
\begin{frontmatter}



\title{Towards Geometric-Photometric Joint Alignment for Facial Mesh Registration}

%

\author{Xizhi Wang\corref{cor1}\fnref{zjlab}}
\ead{yvonnewxz826@163.com}
\author{Yaxiong Wang\fnref{hefei}}
\ead{wangyx15@stu.xjtu.edu.cn}
\author{Mengjian Li\fnref{zjlab}}
\ead{limengjian@zhejianglab.org}
\cortext[cor1]{Corresponding author}
\affiliation[zjlab]{
            organization={Zhejiang Lab},
            addressline={Kechuang Avenue, Yuhang District}, 
            city={Hangzhou},
            postcode={311121}, 
            state={Zhejiang},
            country={China}}
\affiliation[hefei]{
            organization={Hefei University of Technology},
            addressline={Rd.Tunxi No.193}, 
            city={Hefei},
            postcode={230009}, 
            state={Anhui},
            country={China}}

\begin{abstract}
This paper presents a \textbf{G}eometric-\textbf{P}hotometric \textbf{J}oint \textbf{A}lignment~(GPJA) method, which aligns discrete human expressions at pixel-level accuracy by combining geometric and photometric information. Common practices for registering human heads typically involve aligning landmarks with facial template meshes using geometry processing approaches, but often overlook dense pixel-level photometric consistency. This oversight leads to inconsistent texture parametrization across different expressions, hindering the creation of topologically consistent head meshes widely used in movies and games. GPJA overcomes this limitation by leveraging differentiable rendering to align vertices with target expressions, achieving joint alignment in both geometry and photometric appearances automatically, without requiring semantic annotation or pre-aligned meshes for training. It features a holistic rendering alignment mechanism and a multiscale regularized optimization for robust convergence on large deformation. The method utilizes derivatives at vertex positions for supervision and employs a gradient-based algorithm which guarantees smoothness and avoids topological artifacts during the geometry evolution. Experimental results demonstrate faithful alignment under various expressions, surpassing the conventional non-rigid ICP-based methods and the state-of-the-art deep learning based method. In practical, our method generates meshes of the same subject across diverse expressions, all with the same texture parametrization. This consistency benefits face animation, re-parametrization, and other batch operations for face modeling and applications with enhanced efficiency.
\end{abstract}



\begin{keyword}
Geometry registration \sep Facial performance capture \sep Face modeling



\end{keyword}

\end{frontmatter}



\section{Introduction}
\label{sec:introduction}

\begin{figure}[t]
  \centering
  \includegraphics[width=\textwidth]{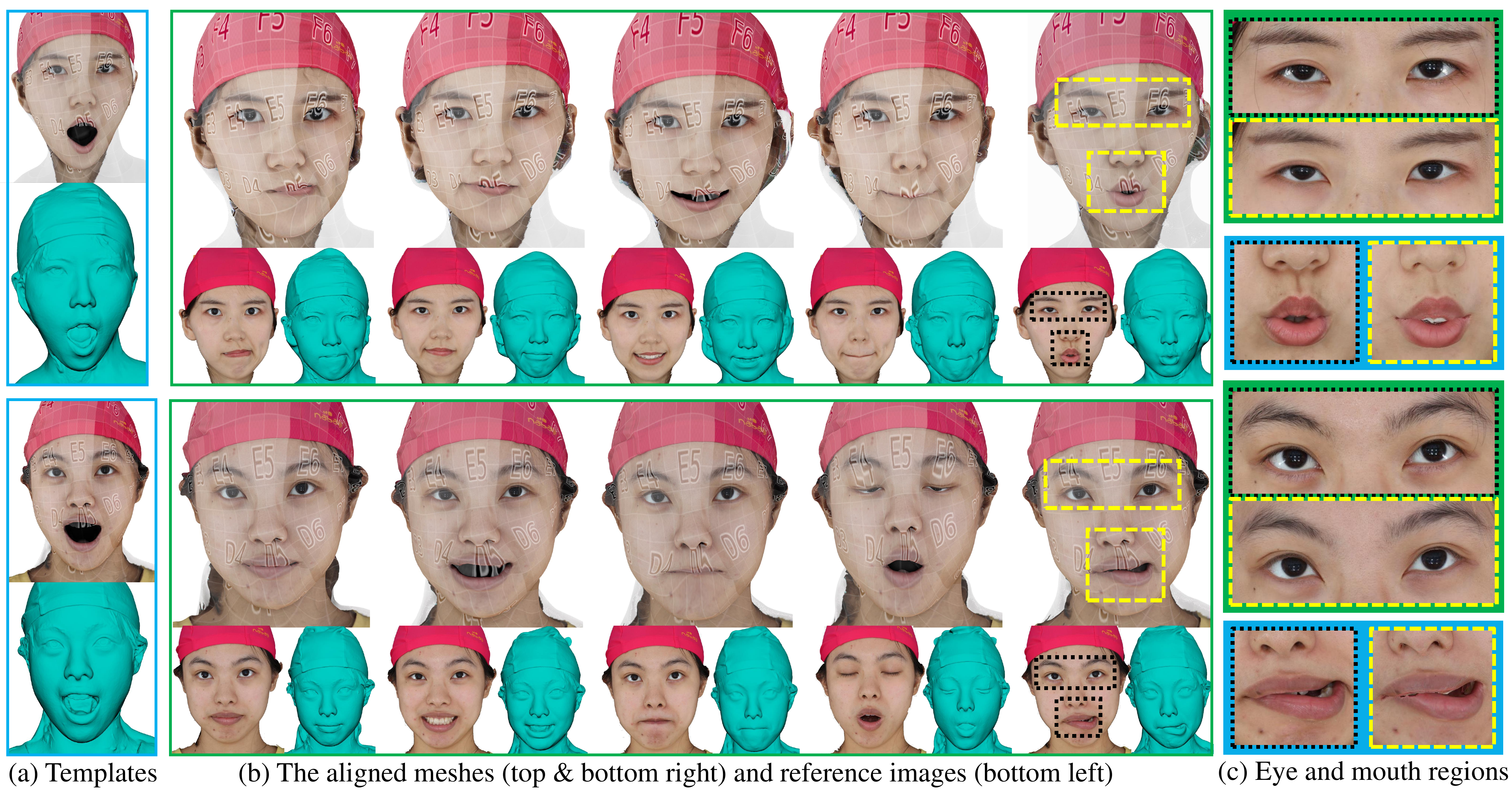}
  \caption{Given multiview images and (a) the textured template mesh, we propose a novel method GPJA based on differentiable rendering to achieve geometric and photometric alignment jointly for facial meshes. (b) The aligned meshes are rendered with the shared texture map as the template in (a). (c) The zoomed renderings~(yellow boxes) of eyes and mouths demonstrate photometric alignment with the reference images~(black boxes).
  }
  \label{fig:teaser}
\end{figure}

Nowadays, professional studios in industry and academia commonly use synchronized multiview stereo setups for facial scanning~\cite{GotardoRBGB18,RiviereGBGB20,ZhangZZLCYXY22}, ensuring high-fidelity results in controlled settings. These setups aim to generate topology-consistent meshes for different subjects with various facial expressions. Typically, conventional pipelines~\cite{FyffeNHSBJLD17} involve constructing raw scans from multiview images, followed by manual processes like marker point tracking, clean-up, or key-framing~\cite{BeelerHBBBGSG11}, which is labor-intensive and time-consuming, limiting its application in film, gaming, AR/VR industry. To fulfill automatic registration, geometry-based methods have been widely employed~\cite{AmbergRV07,Flame17,GilaniMSR18,LeeK19,FanPX23}. However, these methods primarily focus on geometric alignment but failing to ensure photometric consistency at dense pixel-level. To remedy this issue, this paper aims to achieve a \textbf{joint alignment} in terms of geometry and photometric appearances. To this end, two challenges need to be addressed. 

The first challenge is to establish a proper deformation field to guide the alignment process, especially for the challenging areas such as mouths and eyes. Previous attempts have been made to construct correspondences by landmarks or optical flow to aid the photometric alignment~\cite{FyffeNHSBJLD17,AmbergRV07}. However, the offset vectors obtained through these methods often introduce errors. Moreover, since they are extracted from 2D images, these vectors are insufficient for guiding deformation of 3D geometry. ~\cite{GafniTZN21} employed implicit volumetric representation to combine shape and appearance recovery for realistic rendering, which lacks explicit geometry constraints. In response to the challenge, we propose a differentiable rendering~\cite{LaineHKSLA20,Nimier-DavidVZJ19} based registration framework to generate topology-consistent facial meshes from multiview images. In particular, our approach includes a Holistic Rendering Alignment (HRA) which incorporates constraints from color, depth and surface normals, facilitating alignment through automatic differentiation without explicit correspondence computation.

The second challenge for facial mesh registration is generating faithful output meshes while preserving the topological structure. Aligning discrete facial expressions involves large-step geometry deformation, which is susceptible to topological artifacts. Previous works addressed this with specialized constraints~\cite{GilaniMSR18,FanPX23}, but at the cost of a complex pipeline. Other approaches~\cite{BlanzV99,AldrianS13,Flame17,PloumpisVSMWPSG21} utilize parametric models to fit the facial images for mesh creation. Although these methods are highly efficient, they struggle with diverse identities and complex expressions due to their limited expressive capabilities. To overcome this, we resort to a multiscale regularized optimization that combines a modified gradient descent algorithm~\cite{NicoletJJ21}, with coarse-to-fine remeshing scheme. Starting with the coarsest template, the mesh is tessellated periodically while updating the vertices with a robust regularized geometry optimization for the constraints collected from HRA. Our multiscale regularized approach is simple to implement and ensures smoothness and robust convergence, without the need for extensive training data.

We validate our method through experiments on seven subjects from the FaceScape~\cite{Yang0WHSYC20} dataset, covering diverse facial expressions, as shown in Fig.~\ref{fig:teaser}. The joint alignment is examined by both geometric and image metrics, demonstrating the effectiveness of our approach. The aligned meshes produced by our method are of high quality, free from topological errors, and accurately warped even in challenging regions like mouths and eyes.

Our contributions are summarized as following: 
\begin{itemize}
\item A simple and novel method named GPJA achieving joint alignment in geometry and photometric appearances at dense pixel-level for facial meshes.
\item A holistic rendering alignment mechanism based on differentiable rendering that effectively generates the deformation field for joint alignment, without any semantic annotation.
\item A multiscale regularized optimization strategy that ensures robust convergence and produces high-quality, topologically consistent aligned meshes.
\end{itemize}

\section{Related Work}
\label{sec:relwork}
Our research focuses on the registration of facial meshes for topology-consistent geometry on discrete expressions. This section provides a literature review relevant to our study.

\noindent\textbf{Geometry Processing Methods.} Non-rigid registration is a well established technique in geometry processing for warping a template mesh to raw scans~\cite{SorkineCLARS04,RusinkiewiczL01,BadenCK18,cg1_ICP_reg}. The Iterative Closest Point (ICP) algorithm is a commonly used framework~\cite{AmbergRV07,RusinkiewiczL01,LiSP08} for registration. With a template based on 3D Morphable Models (3DMMs) of strong geometric priors~\cite{AmbergRV07,booth20163d,Flame17,GilaniMSR18,LeeK19,Yang0WHSYC20,DaiPSD20,FanPX23} as initialization, ICP minimizes the error between landmarks on the template and the scans, resulting in a rough alignment. Then, a fine-tuning stage involves searching for valid correspondences in the spatial neighborhood, and warping the template leveraging data fidelity and smoothness terms. Previous works have explored regularization terms~\cite{WuBGB16,PearsDSS23} and correspondences~\cite{TamCLLLMMSR13,cg2_ICP_reg} in ICP-based algorithms. In addition to registration, there are also face tracking methods~\cite{bouaziz2013online,seol2016creating} that rely on coefficient regression of rigged blendshapes, which are primarily used for expression transfer rather than high-fidelity reconstruction. Overall, these methods are limited in achieving pixel-level photometric consistency, where our method is successfully capable of.

To achieve photometric alignment, industry-standard methods often involve re-topologizing frame-by-frame using the professional software like Wrap4D~\cite{topo4D_24}, which requires significant time and resources~\cite{DinevBBBXK18}. For automatic pipelines, researchers have explored incorporating optical flow~\cite{cg3_ICP_reg_color,PradaKCCH16,CaoBZB15} into the fine-tuning stage of facial mesh registration. However, optical flow alone is inadequate for handling significant differences between the source template and target scans, as well as occlusion changes around the eyes and mouth~\cite{0001ZWBPBT16}. It is even worse for discrete expression scenarios due to the lack of temporal coherence and large deformation. In contrast, our method effectively handles significant deformation, visibility changes and asymmetry.

Another challenge in facial mesh registration is maintaining smooth contours in facial features~\cite{BermanoBKBBG15}, which is difficult due to occlusions and color changes. Previous approaches have used user-guided methods~\cite{DinevBBBXK18} or contour extraction~\cite{0001ZWBPBT16} to address these challenges, but they either involve lengthy workflows or specific treatments, which hinder efficient topology-consistent mesh creation. On the other hand, our method achieves pixel-level alignment without any semantic annotation.

\noindent\textbf{Deep Learning Based Methods.} Recent advancements in deep learning approaches for generating topology-consistent meshes include ToFu~\cite{LiLBL0Z21}, which predicts probabilistic distributions of vertices on the face mesh to reconstruct registered face geometry. TEMPEH~\cite{BolkartLB23} enhances ToFu with a transformer-based architecture, while NPHM~\cite{GiebenhainKGRAN23} models head geometry using a neural field representation that parametrizes shape and expressions in disentangled latent spaces. Although these methods prioritize geometric alignment, they do not guarantee rigorous photometric consistency. ReFA~\cite{LiuCCZZ22} introduces a recurrent network operating in the texture space for predicting positions and texture maps. Although these deep learning methods represent progress in facial mesh registration, they require a substantial amount of registered data processed with classical ICP-based algorithms. Besides, out-of-domain expressions may be limited by insufficient training data. In our work, no additional training data is needed except for a semi-automatic created template mesh with textures.

In addition to mesh-based representations, implicit volumetric representations have gained popularity in reconstruction. Several studies have extended NeRF~\cite{MildenhallSTBRN20} and 3D Gaussian Splatting~\cite{GS2023} for dynamic face reconstruction~\cite{PumarolaCPM21,ZhengABCBH22,AtharXSSS22,xu2024gaussian,qian2024gaussianavatars}. The pipeline typically begins with explicit parametric models, followed by estimating a deformation field represented by multi-layer perceptrons. Finally, a volumetric renderer is used to generate densities and colors. However, volumetric rendering-based methods lack supervision for aligning mesh-represented geometry and generally do not produce production-ready geometries despite decent rendering results~\cite{LiuCCZZ22}.

\section{Preliminaries}
\label{sec:Preliminaries}
Before discussing the details of our methodology, we first introduce differentiable rendering as background information to provide context for our approach.

Given a 3D scene containing mesh-based geometries, lights, materials, textures, cameras etc., a renderer synthesizing a 2D image \textbf{\textit{P}} of each screen pixel (\textit{x},\textit{y}) can be formulated as:
\begin{equation}
  \boldsymbol{P}(x,y) = F(\boldsymbol{x}|\Theta ),
\end{equation}
where the function $F(\cdot)$ represents the rendering process, encompassing various computations such as shading, interpolation, projection, and anti-aliasing. The output  \textbf{\textit{P}} of this function can be RGB colors, normals, depths, or label images. In our scenario, the parameter to be optimized is the positions of mesh vertices denoted by $\boldsymbol{x}\in \mathbb{R}^{n\times 3}$. $\Theta$ symbolizes a set of scene parameters known in advance, including camera poses, lighting, texture color, and other relevant factors.

Differentiable rendering augments renderers by providing additional derivatives with respect to certain scene parameters, which is a valuable tool for inverse problems~\cite{Nimier-DavidVZJ19,MunkbergCHES0GF22}. The objective of inverse rendering is to recover some specific scene parameters through gradient-based optimization on a scalar loss function $\mathcal{L}$ which is usually defined as the sum of pixel-wise differences between the rendered images $\boldsymbol{P}_{j}$  and the reference image $\boldsymbol{P}_{j}^{\textit{\textbf{r}}}\in \mathbb{R}^{w\times h\times c}$ at the \textit{j}-th camera pose across $v$ views. In this work, we adopt the $L_{1}$-norm for loss functions.
\begin{equation}
  \mathcal{L}\left(\boldsymbol{x}\mid\Theta,\boldsymbol{P}^{\textit{\textbf{r}}}\right )=\sum_{j}^{v}\left | F(\boldsymbol{x}\mid \Theta _{j})- \boldsymbol{P}_{j}^{\textit{\textbf{r}}}\right |.
\end{equation}
In our joint alignment registration setting, the derivative $\frac{\partial \mathcal{L}}{\partial \boldsymbol{x}}$ obtained through differentiable rendering plays a key role, as it guides the template mesh to fit the target expressions. This process avoids the explicit computation of correspondences, such as optical flow, facial landmarks, or other semantic information.

\section{Joint Alignment of Facial Meshes}
\label{sec:method}

\begin{figure*}[t]
  \centering
  \includegraphics[width=\textwidth]{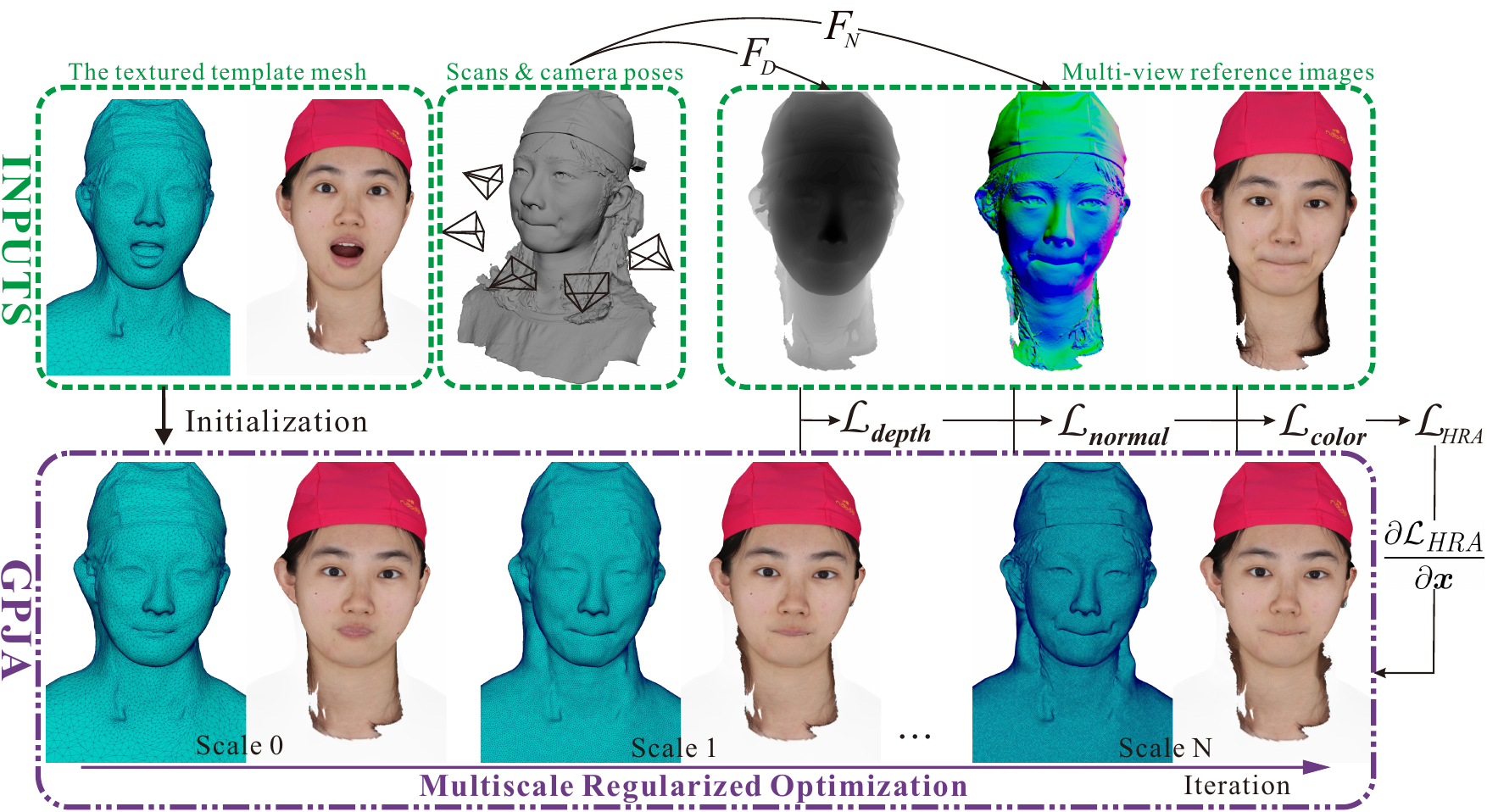}
  \caption{Illustration of our proposed GPJA. With the provided textured template, scans and camera poses, the constraints from HRA $\mathcal{L}$ back-propagates derivatives at each iteration to guide the warping of the template. The regularized optimization is built on a multiscale scheme with periodic tessellation, and the vertices are updated with a robust modified gradient descent algorithm.}
  \label{fig:mainpipeline}
\end{figure*}

Our work addresses the challenge of registering the same subject across diverse facial expressions. Fig.~\ref{fig:mainpipeline} depicts the pipeline of our method. Starting with a textured template mesh and raw scans, each iteration computes deformation from HRA and optimizes the vertices through multiscale regularized optimization. As a result, the outputs of GPJA share a joint alignment across all expressions within the subject.

\subsection{Holistic Rendering Alignment}
\label{ssec:lossfunc}

\begin{figure}
    \centering
    \begin{subfigure}[b]{0.45\textwidth}  
        \centering
        \includegraphics[height=4cm]{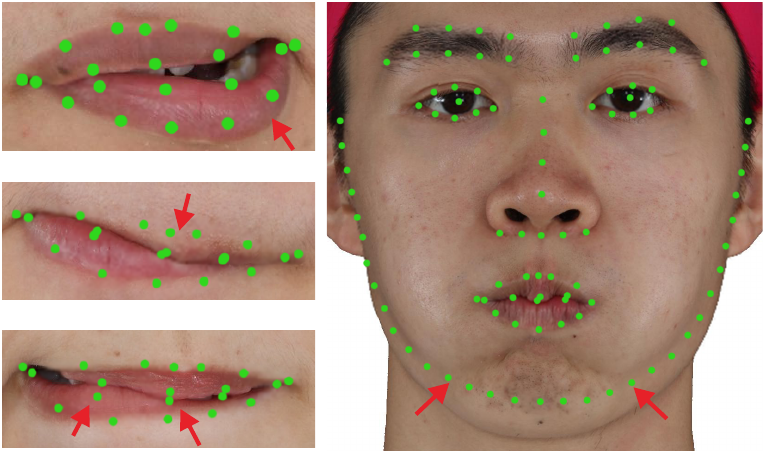}  
        \caption{Common errors with landmarks}  
        \label{fig:image1}
    \end{subfigure}\hspace{0.05\textwidth}  
    \begin{subfigure}[b]{0.45\textwidth}  
        \centering
        \includegraphics[height=4cm]{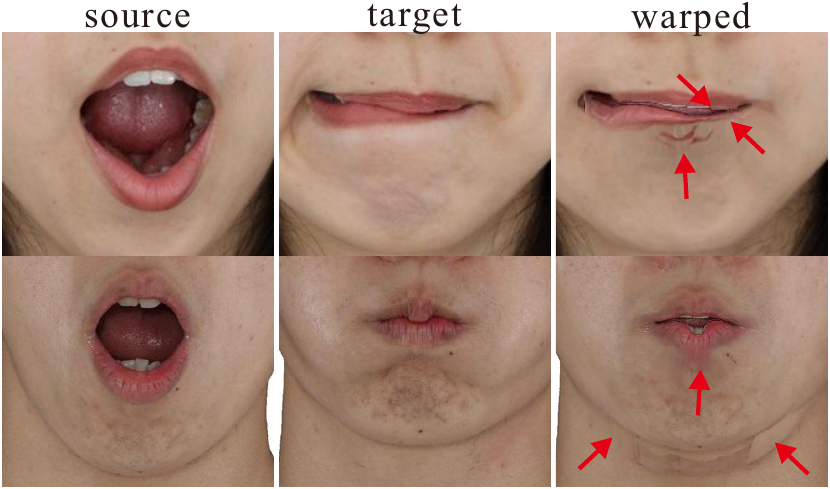}  
        \caption{Common errors with optical flow}  
        \label{fig:image2}
    \end{subfigure}
    \caption{Errors are indicated with arrows: (a) Landmark~\cite{SPIGA} errors under asymmetry, occlusion, and extreme expressions, and (b) Optical flow~\cite{liu2009beyond} inaccuracies due to significant visibility changes.} 
    \label{fig:lmk_OF_error}
\end{figure}

\noindent\textbf{Motivation.} While geometric alignment has been well studied, previous photometric alignment approaches mainly consider sparse facial landmarks, resulting in two main limitations: coarse alignment lacking fine details and unreliability under occlusions and extreme expressions, as shown in Fig.~\ref{fig:lmk_OF_error}(a). Other photometric algorithms~\cite{PradaKCCH16,FyffeNHSBJLD17} attempt to learn deformation from optical flow obtained in the 2D images, but are prone to false correspondences as shown in Fig.~\ref{fig:lmk_OF_error}(b) ,and suffer from ambiguities when elevating to the 3D space. 

Instead of making remedies for landmarks or optical flows, we leverage differentiable rendering, which automatically computes derivatives $\frac{\partial \mathcal{L}} 
{\partial \boldsymbol{x}}$ to guide the deformation of the template mesh ${\boldsymbol{T}}$ into both geometric and photometric alignment. This approach eliminates correspondence errors from facial landmarks~\cite{AmbergRV07} or optical flow offset vectors~\cite{PradaKCCH16}. Furthermore, differentiable rendering operates directly in 3D space, effectively handling occlusions in the eye and mouth regions.

With the above consideration, we propose a Holistic Rendering Alignment (HRA) mechanism, incorporating multiple cues with the aid of differentiable rendering. HRA collects constraints from three different aspects, \emph{i.e.,} color, depth, and surface normals:
\begin{equation}
 \mathcal{L}_{HRA} =  \mathcal{L}_{color} +  \mathcal{L}_{depth} + \mathcal{L}_{normal}.
\end{equation}

\noindent\textbf{Color Constraint.} $\mathcal{L}_{color}$ seeks to impose constraints for photometric alignment by comparing the rendered image with the observed multiview color images. In practice, we found that the deformation around the inner lip is sometimes affected by occlusion changes, resulting in inaccurate lip contours. To address this issue, we deliberately exclude the interior mouth from the color constraint by a masking strategy.

\begin{figure}
  \centering
   \includegraphics[width=0.6\textwidth]{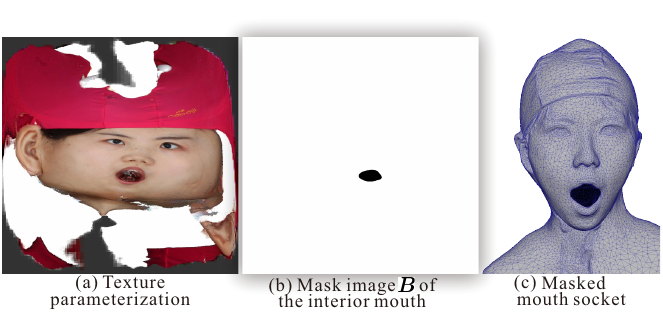}
   \caption{Illustration of the masking strategy. Using the same parametrization as (a) the texture map, the interior mouth is manually masked out, resulting in (b) a binary mask image. By the shading operation $F_{S}(\boldsymbol{x}|\boldsymbol{P}_{j},\boldsymbol{B})$, (c) the mouth socket is mask out from the color constraint.}
   \label{fig:maskop}
\end{figure}

As illustrated in Fig.~\ref{fig:maskop}, a binary mask image $\boldsymbol{B}\in \left \{ 0,1 \right \}^{t\times t\times 3}$ is manually created in accordance with the color texture $\boldsymbol{C}_{\boldsymbol{T}}\in \mathcal{\mathbb{R}}^{t\times t\times3}$ of the given template ${\boldsymbol{T}}$. The mask image labels the interior mouth region, which corresponds to the mouth socket of $\boldsymbol{T}$.

The color constraint is defined as the summation of the element-wise absolute difference between the rendered image and the reference image, weighted by the mask.
\begin{equation}
\label{eq:colorloss}
\begin{split}
\mathcal{L}_{color}\left ( \boldsymbol{x} \right ) = \sum _{j}^{v}| \left (  F_{S}\left (  \boldsymbol{x}\mid  \boldsymbol{\Pi}_{j},\boldsymbol{S},\boldsymbol{C}_{\boldsymbol{T}} \right )-\boldsymbol{P}_{j}^{\textit{\textbf{r}}} \right ) \odot 
 F_{S}\left ( \boldsymbol{x}\mid \boldsymbol{\Pi}_{j},\boldsymbol{B}  \right ) |,
\end{split}     
\end{equation}
where the shading function $F_{S}:\mathbb{R}^{n\times 3}\rightarrow \mathbb{R}^{w\times h\times 3}$ is defined as rendering diffuse objects with the lighting estimated from the capture setup in spherical harmonics forms $\boldsymbol{S}\in \mathcal{\mathbb{R}}^{9\times 3}$~\cite{RamamoorthiH01a}, $\boldsymbol{\Pi}_{j}\in \mathbb{PL}\left ( 3 \right )$ remarks projective matrix of the $j$-th camera, and $\odot$ denotes element-wise multiplication. By synthesizing a binary image using $F_{S}(\boldsymbol{x}|\boldsymbol{\Pi}_{j},\boldsymbol{B})$, where the interior mouth is assigned zeros and the rest with ones, Equation \ref{eq:colorloss} can mask out the interior mouth for color constraint for faithful lip contours. Experimental results confirm the effectiveness of this mask operation.

\noindent\textbf{Depth Constraint.} $\mathcal{L}_{depth}$ pursues to achieve geometric alignment. The color images observed in real-world scenes exhibit inconsistencies due to variations in shading and imaging formulation across views and expressions. Therefore, the derivatives of color constraint inevitably introduce bias and noise, making disturbances for accurate alignment. By including the depth term in HRA, we provide strong supervision for preserving geometric fidelity during registration, ensuring robust outputs.

The depth constraint measures the depth disparity between the deformed template and the target scan:
\begin{equation}
\mathcal{L}_{depth}\left ( \boldsymbol{x} \right ) = \sum _{j}^{v}| \left (  F_{D}\left (  \boldsymbol{x}\mid  \boldsymbol{\Pi}_{j}\right )- F_{D}\left (  \boldsymbol{\tilde{x}}\mid  \boldsymbol{\Pi}_{j}\right ) \right )  |,  
\end{equation}
where $F_{D}:\mathbb{R}^{n\times 3}\rightarrow \mathbb{R}^{w\times h}$ represents the depth rendering operation\cite{Harlteybible} based on the perspective projection function, and $\boldsymbol{\tilde{x}}\in \mathbb{R}^{m\times 3}$ denotes vertex coordinates of the scan with $m$ vertices. 

\noindent\textbf{Normal Constraint.} $\mathcal{L}_{normal}$ assists with fidelity preservation. While the color and depth terms establish guidance for overall alignment, discrepancies in color tones and texture-less areas can cause artifacts on aligned meshes. The normal constraint compensates for these issues, leading to improved alignment and sharper details with fewer vertices.

In specific, the normal constraint penalties the disparity of surface normals between the deformed template and the target scan.
\begin{equation}
\mathcal{L}_{normal}\left ( \boldsymbol{x} \right ) = \sum _{j}^{v}| \left (  F_{N}\left (  \boldsymbol{x}\mid  \boldsymbol{\Pi}_{j}\right )- F_{N}\left (  \boldsymbol{\tilde{x}}\mid  \boldsymbol{\Pi}_{j}\right ) \right )  |,
\end{equation}
where $F_{N}:\mathbb{R}^{n\times 3}\rightarrow \mathbb{R}^{w\times h\times 3}$ represents the process of computing and projecting surface normals, as implemented in deferred shading~\cite{GpuGems3}.

HRA benefits from differentiable rendering to obtain derivatives, replacing the explicitly computed correspondences from previous approaches~\cite{TamCLLLMMSR13,DengYDZ22}. By incorporating multiple cues, HRA ensures joint alignment of geometry and photometric appearances.

\subsection{Multiscale Regularized Optimization}
\label{ssec:multiscaleopt}
The HRA mechanism steers deformation towards joint alignment, yet its derivative vectors are noisy because of shading and imaging variations. Since the derivatives lack regularization, applying them directly as update steps to each vertex could lead to topological errors~\cite{NicoletJJ21,JungKHB023}. A common way to tackle this is adding a regularization term~\cite{TamCLLLMMSR13,DengYDZ22}, like the ARAP energy~\cite{SorkineA07} or the Laplacian differential representation \cite{SorkineCLARS04}. However, these solutions introduce problems in tuning the regularization weight for outputs with both smooth and non-smooth regions~\cite{NicoletJJ21}, and implementing a robust solution scheme for non-linear optimization~\cite{SchmidtBBCK22}. To overcome these challenges, we propose a multiscale regularized optimization for generating high-quality aligned meshes.

\noindent\textbf{Vertex Optimization.} We follow the work of \cite{NicoletJJ21} to update vertices iteratively. It suggests that the second-order optimization like Newton's method is better for smoothing geometry, and the computationally expensive Hessian matrix can be replaced by re-parametrization of $\boldsymbol{x}$ with the introduced variables $\boldsymbol{\mu }$ to ensure the smoothness of recovered $\boldsymbol{x}$: 
\begin{equation}
\label{eq:u}
\boldsymbol{x}= \left( \boldsymbol{I}+ \lambda \boldsymbol{L}\right)^{-1} \boldsymbol{\mu }, 
\end{equation}
\begin{equation}
\label{eq:x}
\boldsymbol{\mu }\leftarrow \boldsymbol{\mu }- \eta \frac{\partial \boldsymbol{x}}{\partial \boldsymbol{\mu }}\frac{\partial \mathcal{L} }{\partial \boldsymbol{x}},
\end{equation}
where $\eta> 0$ means the learning rate, $\boldsymbol{I} \in \mathbb{I}^{n\times n}$ denotes identity matrix, and $\lambda> 0$ is the regularization weight. $\boldsymbol{L} \in \mathbb{R}^{n\times n}$ is a discrete Laplacian operator defined on a mesh  $\mathcal{M}= \left (  \mathcal{X}, \mathcal{E}\right )$ with $n$ vertices $\mathcal{X}$ and $m$ edges $\mathcal{E}$:
\begin{equation}
\boldsymbol{L}_{ij}= \left\{\begin{matrix}
-w_{ij}, & \textrm{if}\quad(i,j) \in \mathcal{E}\\ 
\sum _{(i,k)\in\mathcal{E}}w_{ik} & \mathrm{if}\quad i =j \\ 
 0,& \mathrm{otherwise}, 
\end{matrix}\right.\\
\end{equation}
where $w_{ij}$ is the cotangent weight described in \cite{DesbrunMSB99}.

Combining Equations \ref{eq:u} and \ref{eq:x}, the updated formula for vertices at each iteration is:
\begin{equation}
\boldsymbol{x}\leftarrow \boldsymbol{x}-\eta \left ( \boldsymbol{I}+  \lambda \boldsymbol{L}\right)^{-2}\frac{\partial \mathcal{L}_{HRA} }{\partial \boldsymbol{x}}.
\end{equation}

\noindent\textbf{Multiscale Learning.} The coarse-to-fine multiscale learning scheme, based on the tessellation technique~\cite{BotschK04}, periodically decreases the average edge length of the triangle mesh while retaining the shape and the texture parametrization. The multiscale scheme allows for parameter adjustment at each scale to capture fine details without distorting the topology. In particular, the Laplacian matrix $\boldsymbol{L}$ is updated for each tessellation step as the topology changes. The template mesh $\boldsymbol{T}$, initially at the coarsest level, undergoes more rigid deformations with a higher regularization parameter $\lambda$ to fit the overall target expressions. As the mesh is tessellated to finer scales, $\lambda$ is decreased to capture more details.

The multiscale regularized optimization offers several advantages: (1) it produces high-quality meshes effectively with significantly reduced distortion and self-intersection artifacts; (2) it converges robustly without additional training data or priors other than the textured template. The iterative optimization process can be observed in the supplementary videos.

\begin{figure}[t]
  \centering
  \includegraphics[width=\textwidth]{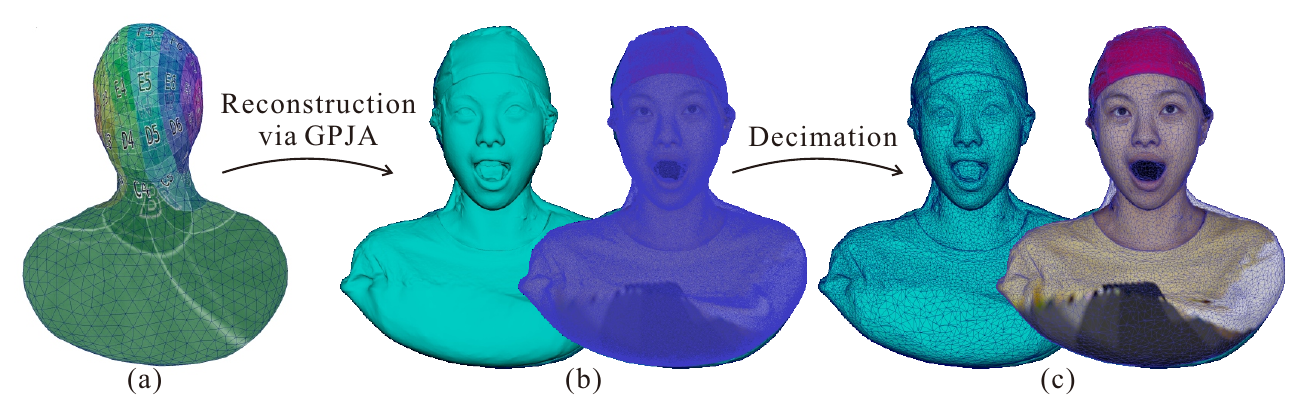}
  \caption{The pipeline of the textured template mesh creation. (a) A genus-0 mesh in the shape of a bust that approximately overlaps the scan undergoes GPJA as Fig.~\ref{fig:mainpipeline}, producing (b) a densely tessellated mesh accompanied with the reconstructed texture map $\boldsymbol{C}_{\boldsymbol{T}}$, which is then decimated to construct (c) a coarse template $\boldsymbol{T}$, while still preserving the same texture map.}
  \label{fig:templatecreat}
\end{figure}

\subsection{Textured Template Mesh Creation}
\label{ssec:templatecreate}
To adapt to the proposed pipeline, we construct a reliable textured template mesh which is used throughout the joint alignment for different expressions per subject. As illustrated in Fig.~\ref{fig:templatecreat}, we deliberately choose the mouth-open expression from each subject to reconstruct it into the template mesh. This is because the mouth socket is crucial for accommodating flexible movements during deformation. Initially, we manually sculpt a coarse genus-0 mesh with pre-designed texture parametrization, resembling a bust sculpture. This genus-0 surface is appropriate for representing the head geometry due to similar topological structures without loss of generality .

The genus-0 bust mesh as initialization is then processed through the GPJA pipeline. At this stage, only the depth constraint in HRA is applied. After the depth guiding reconstruction, a tessellated mesh is established. The geometry is then fixed, and the texture color map $\boldsymbol{C}_{\boldsymbol{T}}$ is updated through the color constraint via gradient descent, shown in Fig.~\ref{fig:templatecreat}(b). Finally, the tessellated mesh is decimated using seam-aware simplification~\cite{LiuFJG17} while preserving the texture parametrization, resulting in the creation of the textured template mesh $\boldsymbol{T}$ as the startup mesh in Fig.~\ref{fig:mainpipeline}. 

Although the template is subject-specific, all expressions share a common texture space, which  enables batch re-parametrization. This offers a practical way for topology-consistent meshes of large datasets.

\section{Experiments and Analysis}
\label{sec:results}
\noindent\textbf{Experiment Setup.} As an early exploration of semantic annotation-free photometric alignment, many existing public datasets (LYHM~\cite{DaiPSD20}, and NPHM~\cite{GiebenhainKGRAN23}, etc.) that primarily consist of 3D scans or registered meshes rather than original images are unsuitable for GPJA. We also found FaMoS~\cite{BolkartLB23} inappropriate due to its sparse down-sampled RGB views and subjects with noticeable facial markers. Following our investigation, the FaceScape dataset emerged as the most fitting benchmark with high-resolution images from dense viewpoints and uniform lighting conditions for discrete facial expressions.

In order to thoroughly assess GPJA's capability, seven subjects, including four publishable ones~(Subject 122, 212, 340 and 344), are chosen, and we deliberately selected 10 highly different expressions for each subject. The selection process eliminates redundant and similar expressions. Additionally, we prioritize expressions with significant deformation and occlusion compared to the neutral expression, where landmark detection is less accurate as shown in Fig.~\ref{fig:lmk_OF_error}. This process ensures that the chosen expressions cover a range of challenging scenarios and effectively evaluate the algorithm’s performance. Six to eight images, covering frontal and side views, are used as references.

GPJA is implemented with Nvdiffrast~\cite{LaineHKSLA20} as the differentiable renderer and LargeSteps~\cite{NicoletJJ21} as the optimizer. The rendering uses the Lambertian reflection model under uniform lighting, which is a reasonable simplification of FaceScape's acquisition environment. The multiscale optimization involves remeshing 4 times, increasing the vertex count from 16K to 250K. The learning rate $\eta$ is set to one-tenth of the face length. For the first two levels, the regularization parameter $\lambda$ is set to 200 and 120, while for the remaining levels, it is set to 80 and 50. Gradients of only one constraint are back-propagated per iteration, with three constraints being iterated in rotation. Convergence is achieved within 1500 iterations per expression, taking approximately 15 minutes on a single NVIDIA RTX 3080 graphics card.

\noindent\textbf{Metrics.} As GPJA achieves joint alignment, we evaluate the method's effectiveness from both geometric and photometric perspectives. Geometric alignment is assessed using raw face scans as ground truth, with L1-Chamfer distance and normal consistency reported. The distance is computed via the point-mesh approach to reduce the influence of vertex counts. Moreover, we calculate the F-Score with thresholds of 0.5mm and 1.0mm for a statistical error analysis. For photometric alignment, multiview reference images are used as ground truth, while rendered images are generated from aligned meshes with the common template's texture map at the same camera poses. Standard image metrics such as PSNR, SSIM, and LPIPS are employed for evaluation.

\begin{table*}
\caption{Geometric evaluations on 10 expressions per subject in millimeter. C.D. stands for L1-Chamfer distance, N.C. represents normal consistency, while F@0.5 and F@1.0 are F-Scores at thresholds of 0.5mm and 1.0mm.}
\label{tab:geoMetric}
\setlength{\tabcolsep}{1pt} 
\resizebox{\textwidth}{!}{ 
\begin{tabular}{c  |c|c |c|c |c|c |c|c |c|c |c|c |c|c }
\hline
\multirow{2}{*}{\textbf{}} &  \multicolumn{2}{c|}{\textbf{Subject 7}}&  \multicolumn{2}{c|}{\textbf{Subject 32}} & \multicolumn{2}{c|}{\textbf{Subject 122}} & \multicolumn{2}{c|}{\textbf{Subject 212}} & \multicolumn{2}{c|}{\textbf{Subject 340}} & \multicolumn{2}{c|}{\textbf{Subject 344}} & \multicolumn{2}{c}{\textbf{Subject 350}} \\
\cline{2-15}
 & C.D.$\downarrow$ & N.C.$\uparrow$ & C.D.$\downarrow$ & N.C.$\uparrow$ &  C.D.$\downarrow$ & N.C.$\uparrow$ & C.D.$\downarrow$ & N.C.$\uparrow$ & C.D.$\downarrow$ & N.C.$\uparrow$ 
 & C.D.$\downarrow$ & N.C.$\uparrow$ & C.D.$\downarrow$ & N.C.$\uparrow$\\
 \hline
NICP & 0.477 & 0.910 &  0.355 & 0.942&  0.404 & 0.943 & 0.614 & 0.915 & 0.383 & 0.944 & 0.340 & 0.936 & 0.428 & 0.912 \\
NPHM & 0.347 & 0.958 & 0.377 & 0.970 & 0.325 & 0.968 & 0.429 & 0.949 & 0.285 & 0.973 & 0.338 & 0.957 & 0.489 & 0.937  \\
GPJA & \textbf{0.229} & \textbf{0.981} & \textbf{0.147} & \textbf{0.992} & \textbf{0.264} & \textbf{0.984} & \textbf{0.180} & \textbf{0.980} & \textbf{0.215} & \textbf{0.981} & \textbf{0.224} & \textbf{0.975} & \textbf{0.239} & \textbf{0.970} \\
\hline
 & F@0.5$\uparrow$ & F@1.0$\uparrow$ & F@0.5$\uparrow$& F@1.0$\uparrow$  & F@0.5$\uparrow$ & F@1.0$\uparrow$  & F@0.5$\uparrow$ & F@1.0$\uparrow$  & F@0.5$\uparrow$ & F@1.0$\uparrow$  & F@0.5$\uparrow$ & F@1.0$\uparrow$  & F@0.5$\uparrow$ & F@1.0$\uparrow$ \\
 \hline
NICP & 0.708 &0.852 & 0.800 &0.918 & 0.738 &0.901 & 0.549 &0.799 & 0.743 & 0.913 & 0.792 &0.922 & 0.746 &0.879 \\
NPHM & 0.814 & 0.923 &  0.778 & 0.925 & 0.825 & 0.945 &  0.731 & 0.899 &  0.858 & 0.956 &  0.814 & 0.938 &  0.682 & 0.872 \\
GPJA & \textbf{0.909} & \textbf{0.960} & \textbf{0.957} & \textbf{0.975} & \textbf{0.880} & \textbf{0.959} & \textbf{0.921} & \textbf{0.963} & \textbf{0.912} & \textbf{0.961} & \textbf{0.912} & \textbf{0.960} & \textbf{0.891} & \textbf{0.957} \\
\hline

\end{tabular}
}
\end{table*}

\begin{figure*}
  \centering
  \includegraphics[width=\textwidth]{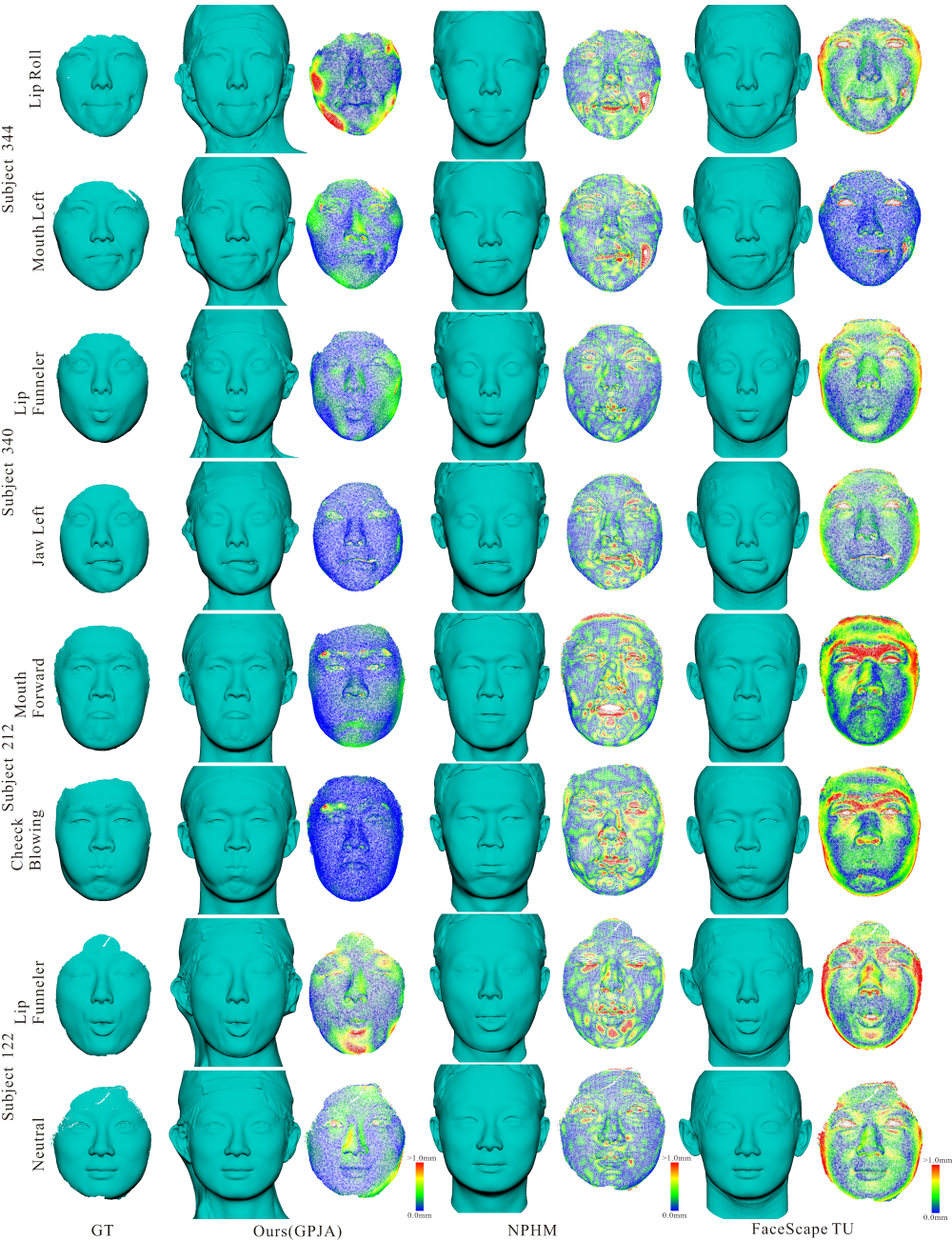}
  \caption{Visualization on the geometric errors. GPJA outputs meshes with higher vertex counts than NPHM and FaceScape TU. To mitigate the effect of vertex counts, we use the point(of GT)-to-mesh distance. The error distribution show that GPJA achieves lower geometric errors compared to NPHM and FaceScape TU.}
  \label{fig:geoerror}
\end{figure*}

\noindent\textbf{Result Analysis on Geometric Alignment.} We compare our GPJA against two registration methods. The first is FaceScape topologically uniform(TU) meshes obtained through NICP~\cite{Yang0WHSYC20} which is a variant of the standard ICP registration method in 3D face domain~\cite{JungKHB023}. The second is the state-of-the-art deep learning method NPHM~\cite{GiebenhainKGRAN23}, which is trained on a dataset comprising 87 subjects and 23 different facial expressions.

Quantitative and qualitative comparisons of geometric alignment are presented in Table~\ref{tab:geoMetric} and Fig.~\ref{fig:geoerror}. Table~\ref{tab:geoMetric} shows that GPJA outperforms NPHM and NICP across all subjects, achieving the lowest L1-Chamfer distance and highest normal consistency, indicating superior geometric alignment. GPJA also consistently achieves higher F-Scores at both thresholds, with F-Scores@0.5mm particularly emphasizing its statistical superiority.

As depicted in Fig.~\ref{fig:geoerror}, the error visualization exhibits the distribution of errors across the face, demonstrating that our method achieves higher fidelity, even in challenging regions such as the lips and eyes. The primary limitation of NPHM is its lack of fidelity in capturing details. This deficiency is particularly evident in errors concentrated around the mouth of subject 212 and wrinkles of subject 344. Fig.~\ref{fig:geoerror} also reveals two main drawbacks in FaceScape TU meshes. First, subject 122's lip-funneler expression, where the eyes should be closed, incorrectly shows the opposite. This is a common issue resulting from inaccurate landmarks in the ICP-based method. Secondly, the face rim areas (foreheads and cheeks) in TU meshes are observed with higher geometric errors due to excessive conformation to the template shape. Additionally, both NPHM and TU meshes incorrectly stitch non-facial areas like the skull, ears, and neck using templates. In contrast, GPJA aligns these regions more accurately, showcasing its robustness in capturing full heads.

\noindent\textbf{Result Analysis on Photometric Alignment.} Previous registration methods neglect photometric evaluations due to a lack of consideration for joint alignment, hence we provide qualitative evaluation for comparisons. 

\begin{figure}
  \centering
  \includegraphics[width=\textwidth]{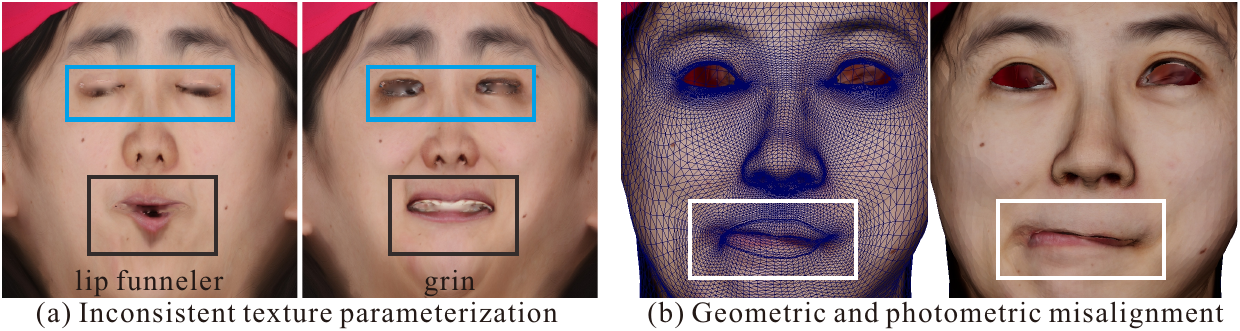}
   \caption{Evidences of photometric misalignment on FaceScape TU meshes.}
  \label{fig:Facescape_inconsis}
\end{figure}

\begin{figure}
  \centering
  \includegraphics[width=\textwidth]{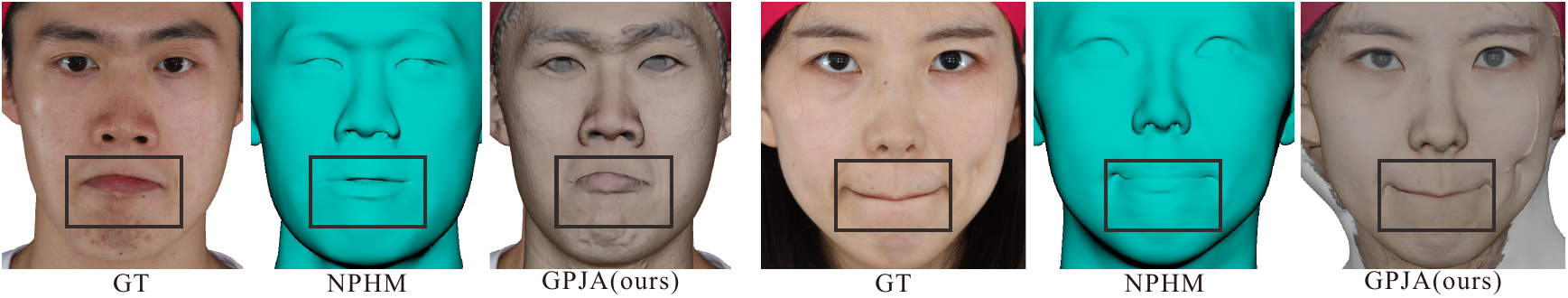}
   \caption{Evidences of photometric misalignment in NPHM. The lip contours on the NPHM mesh do not align with the GT images.}
  \label{fig:NPHM_inconsi}
\end{figure}

There are two key evidences illustrating the lack of photometric alignment in FaceScape, as shown in Fig.~\ref{fig:Facescape_inconsis}: (a) The texture maps for the same subject exhibit discrepancies across different expressions, signifying the absence of a unified parametrization; (b) Deviations are observed between the contours of lips on the texture maps and those on the meshes, revealing misalignment between the geometry and texture. Similar issues are also present in NPHM meshes, although they are not accompanied by texture coordinates. Fig.~\ref{fig:NPHM_inconsi} reveals that NPHM generates false lips that ought to be invisible, which implies the method's insufficient ability to align based solely on geometric features.

\begin{table}
\centering
  \caption{Photometric evaluations on 10 expressions per subject. The image metrics are comparable to the photo-realistic results by the NeRF-style reconstruction NeP~\cite{MaLLWZWS22}.}
  \label{tab:ImgMetric}
\begin{tabular}{l | c c c }
\hline
&\multicolumn{1}{c}{PSNR↑} & \multicolumn{1}{c}{SSIM↑} & \multicolumn{1}{c}{LPIPS↓} \\
\hline
{Subject 7 }     &24.75 & 0.7810 & 0.06868  \\
{Subject 32}     &20.79 & 0.6621 & 0.07516 \\
{Subject 122 }   &23.56 & 0.7443 & 0.07081    \\
{Subject 212 }   &25.51 & 0.7699 & 0.09368   \\
{Subject 340 }   &24.10 & 0.7378 & 0.06802\\
{Subject 344 }   &23.08 & 0.7569 & 0.06791 \\
{Subject 350 }   &25.38 & 0.7320  & 0.07857  \\
\hline
Average & 23.88 & 0.7406 & 0.07469  \\
\hline
\end{tabular}
\end{table}

 \begin{figure*}
  \centering
  \includegraphics[width=\textwidth]{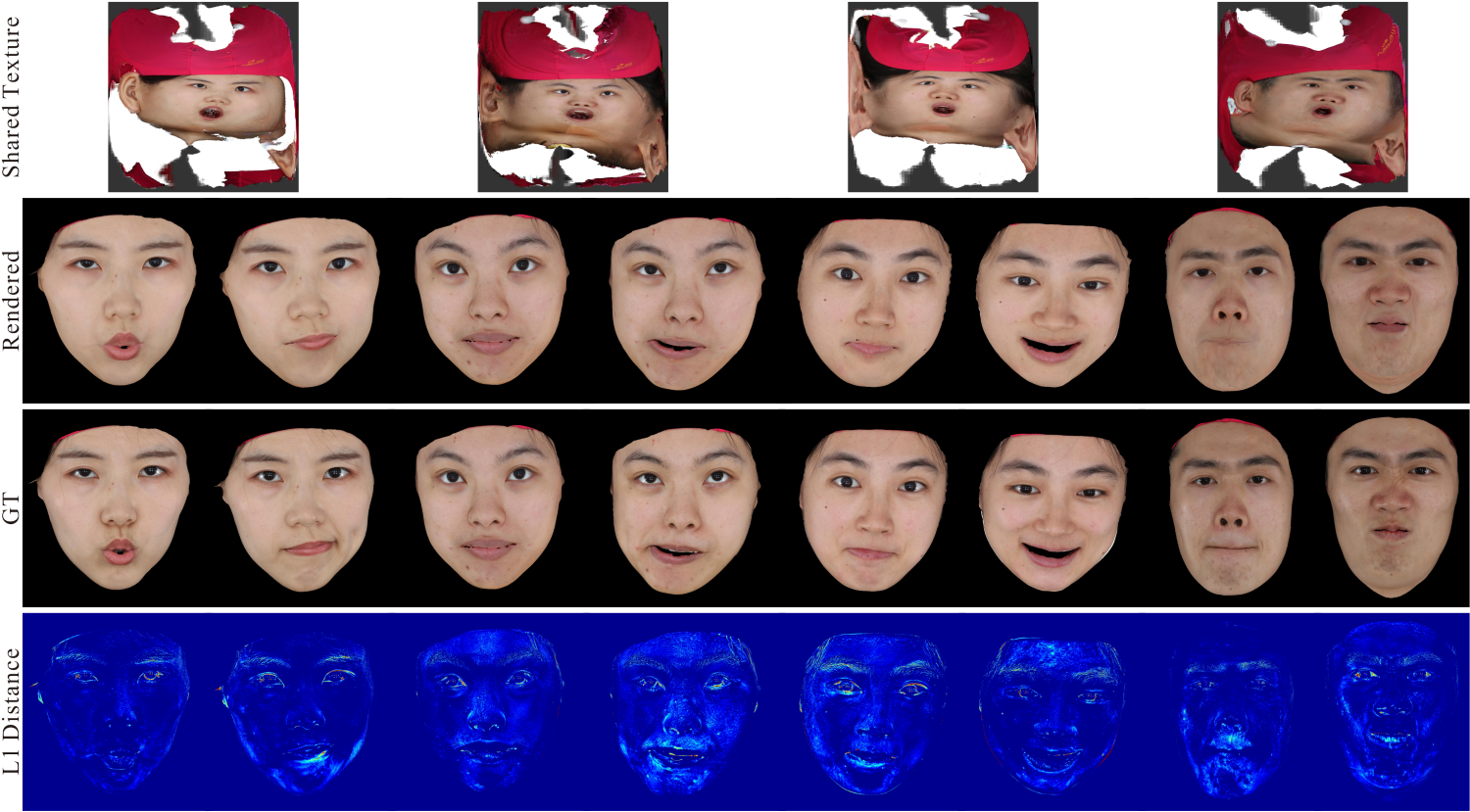}
  \caption{Visualization on photometric errors of GPJA. The expressions of each subject are rendered using the shared color texture shown in the first row, and $L_{1}$ distance is computed by overlapping the rendered images with ground truth.}
  \label{fig:photoerror}
\end{figure*}

\begin{figure}
  \centering
   \includegraphics[width=\textwidth]{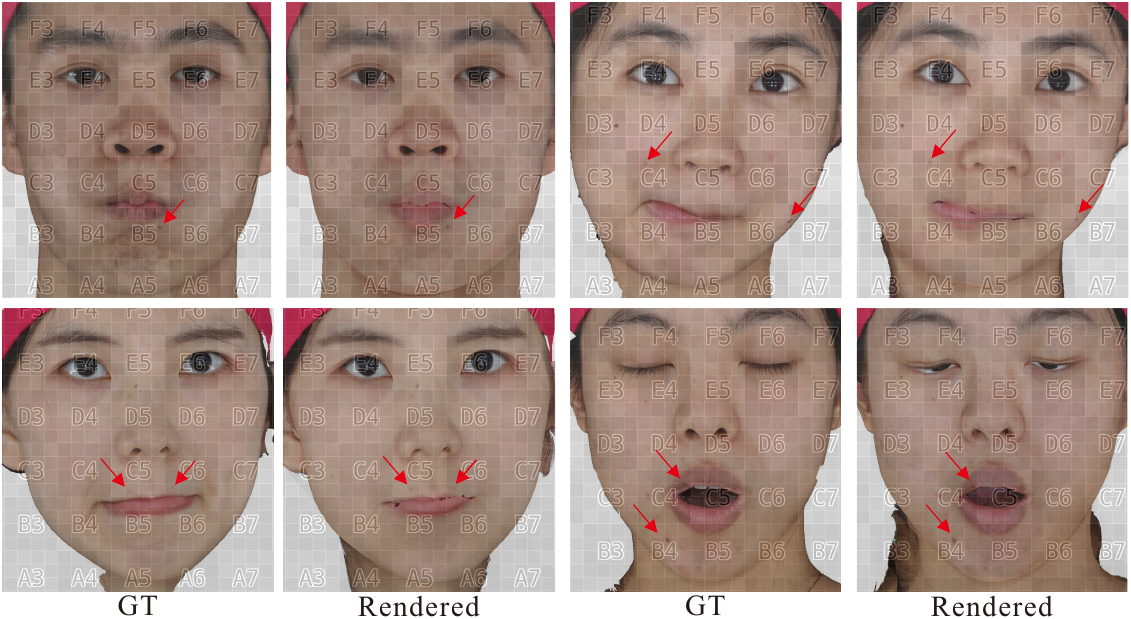}
   \caption{Close-up views of GT and rendered images. Checkerboard patterns are used to demonstrate the photometric consistency of moles and freckles, as indicated by arrows. Zooming in on this figure is recommended.}
   \label{fig:zoomin}
\end{figure}

In contrast, GPJA ensures photometric consistency by applying the shared color texture to match reference images of different expressions. A comparison between the ground truth and our rendered images in Fig.~\ref{fig:photoerror} convincingly shows the photometric alignment achieved by our registration method. The last row of Fig.~\ref{fig:photoerror} illustrates that the discrepancy between the ground truth and the rendered images is primarily attributed to variations in skin tone across facial expressions. Notably, as depicted in Fig.~\ref{fig:zoomin}, the rendered images exhibit pixel-level consistency in characteristic areas such as the mouth, eyes, and mole features across various facial expressions, despite occlusion changes. Moreover, Table~\ref{tab:ImgMetric} provides the image metrics for our experiment. The quantitative results are superior than PSNR of 23.61, SSIM of 0.6460, and LPIPS of 0.09677 achieved by the NeRF-style pipeline NeP~\cite{MaLLWZWS22} (tested on the first 100 subjects of FaceScape), which produces photo-realistic reconstruction through per-frame color generation without alignment. 

\begin{figure}
  \centering
  \includegraphics[width=0.8\textwidth]{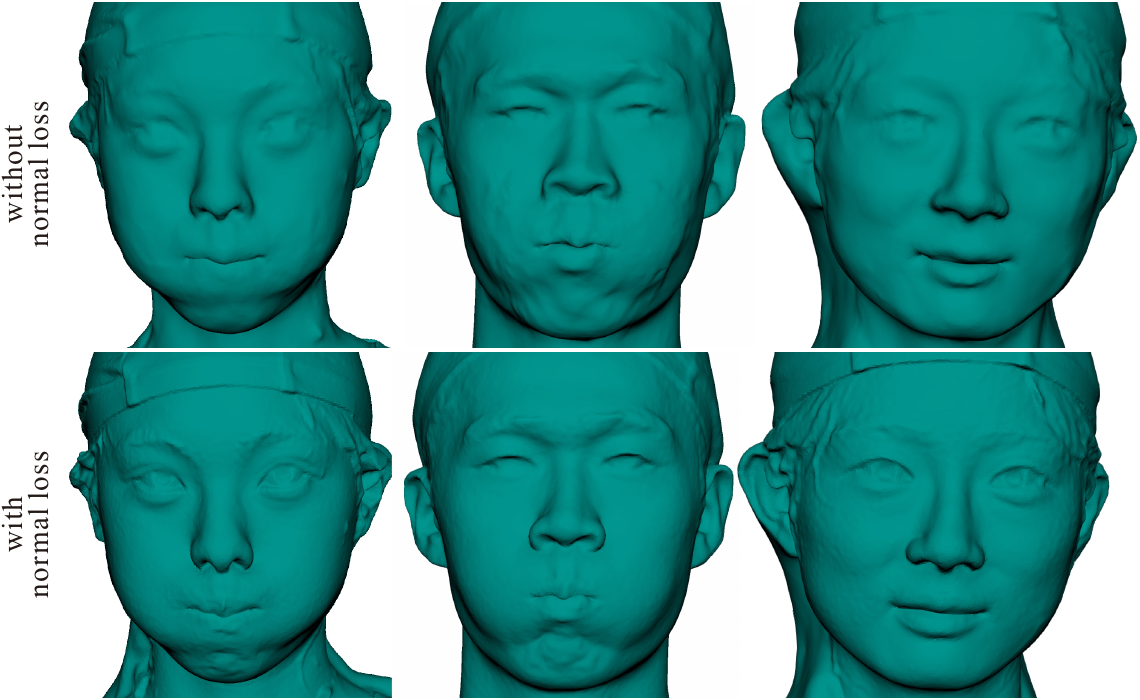}
   \caption{Ablations with and without the normal constraint. The normal constraint in HRA significantly enhanced details. }
  \label{fig:ablanormal}
\end{figure}

\begin{figure}
  \centering
  \includegraphics[width=0.6\textwidth]{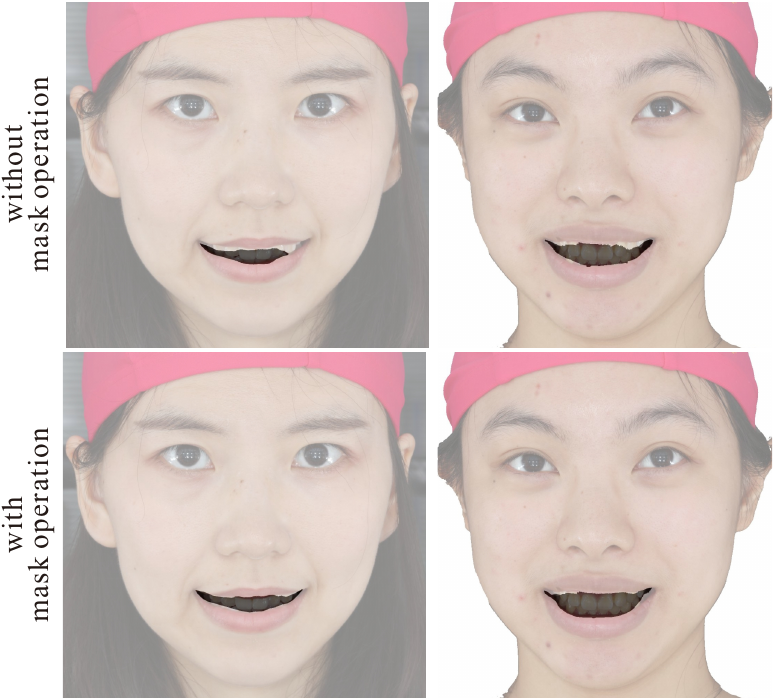}
   \caption{Ablations with and without the mask operation. Masking out the interior mouth avoids disturbance around the contour. }
  \label{fig:ablamask}
\end{figure}

\begin{table*}
\centering
  \caption{Ablation studies demonstrating the effects of each constraint from HRA on four publishable subjects. The geometric error used is the point-to-mesh distance. The geometric performance degrades when any of the three constraints is removed, validating the contributions of each constraint in HRA.}
  \label{tab:alba}
\begin{tabular}{l| c c c c}
\hline
     & Geometric error↓ & {PSNR↑} & {SSIM↑} & {LPIPS↓}\\
\hline
GPJA & \textbf{ 0.167} & 24.06 & 0.7523 & \textbf{0.07511}  \\
\hline
\multicolumn{4}{l}{\textbf{Ablation}}\\ 
\hline
{w/o $\mathcal{L}_{normal}$ } & 0.377 &  \textbf{24.49} & 0.7498 & 0.07578 \\
{w/o $\mathcal{L}_{depth}$ } &  0.455 &   23.67 & \textbf{0.7535}& 0.07544  \\
{w/o $\mathcal{L}_{color}$ } &  0.447 &  22.04 & 0.7364 & 0.07820\\
\hline
\end{tabular}
\end{table*}

\noindent\textbf{Ablations.} The HRA mechanism plays a crucial role in correctly warping the template into joint alignment. To validate HRA's effectiveness, we conduct two ablation experiments to confirm its benefits. 

We first validate the contribution of each constraints from HRA mechanism, which is supported by Table~\ref{tab:alba}. In the case where all constraints are utilized, the geometric error is minimized. As a visualized example, Fig.~\ref{fig:ablanormal} reveals that the normal term significantly contributes to the sharpness of details and alleviates incorrect bumps on texture-less regions like cheeks. However, Table~\ref{tab:alba} show close image metrics in comparison. We speculate the reason is that with the color constraint guiding the deformation, the geometric distortion can not manifest itself in color renderings.

In the second ablation, we study the effectiveness of the masking strategy for color constraint. We synthesize the label image for the mouth socket using $F_{S}(\boldsymbol{x}|\boldsymbol{\Pi}_{j},\boldsymbol{B})$, and overlap it with the ground truth to examine the contours around the inner mouth. Fig.~\ref{fig:ablamask} illustrates that when the inner mouth is not masked out, the vertices around it are disturbed and distorted. Hence, intentionally masking out the inner mouth facilitates correct tracing of the mouth contours.

\noindent\textbf{Applications.} The core strength of GPJA lies in its ability to achieve joint alignment with a shared texture space across expressions. This consistency enables efficient batch processing for a variety of applications, such as animation and re-parametrization, as demonstrated below. 

\begin{figure}
  \centering
  \includegraphics[width=0.9\textwidth]{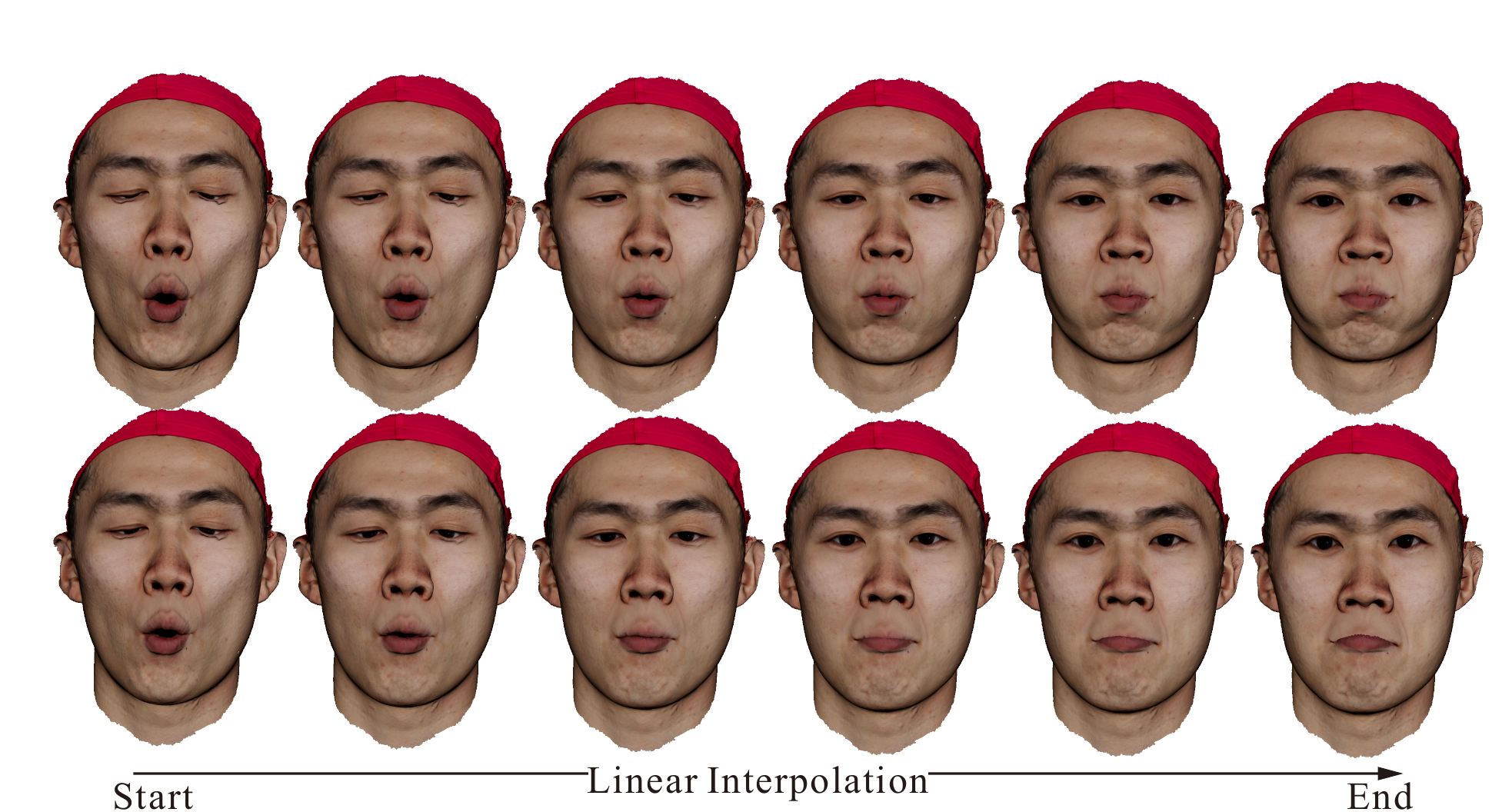}
   \caption{Face animation via vertex interpolation.}
  \label{fig:animation}
\end{figure}

The meshes generated by GPJA for the same subject share a common parametrization, though the vertex sampling varies. We perform remeshing on these meshes via a common triangulation on the texture space. The remeshed results can then be linearly interpolated for face animation as presented in Fig.~\ref{fig:animation} and the supplementary video.

\begin{figure}
    \centering
    \begin{subfigure}[b]{0.3\textwidth}  
        \centering
        \includegraphics[height=7cm]{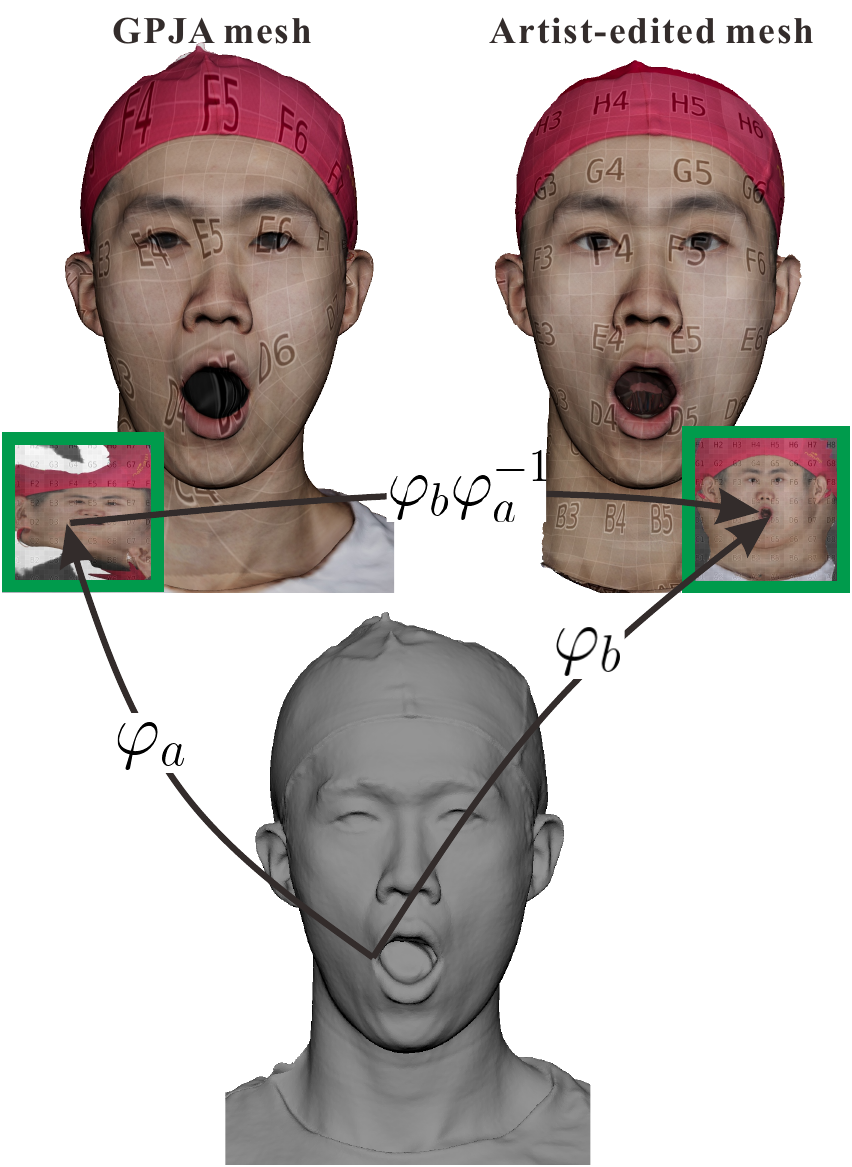}  
        \caption{}  
        \label{fig:image1}
    \end{subfigure}\hspace{0.05\textwidth}  
    \begin{subfigure}[b]{0.6\textwidth}  
        \centering
        \includegraphics[height=7cm]{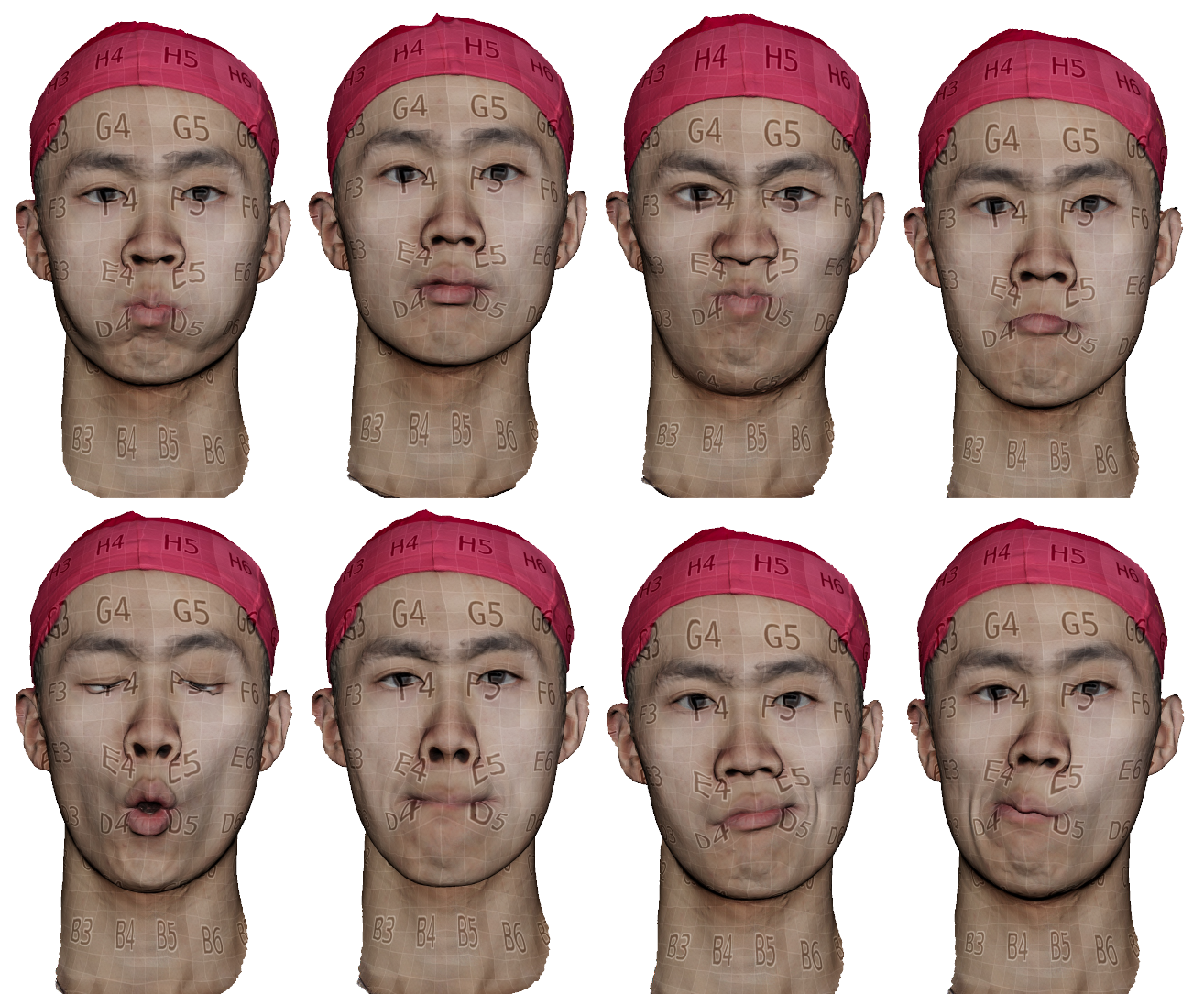}  
        \caption{}  
        \label{fig:image2}
    \end{subfigure}
    \caption{(a) Computation of maps on the texture spaces. (b) Re-parametrization results.}  
    \label{fig:reparam}
\end{figure}

When an artist-edited, UV-unwrapped facial mesh is available, as illustrated in Fig.~\ref{fig:reparam}(a), the texture embeddings of input meshes yield a continuous map $\varphi _{b}\varphi _{a}^{-1}$ from the texture coordinates of the GPJA-generated meshes to those of the artist-edited mesh. Thereafter, the interpolation operator is applied directly to propagate the artist-edited parametrization across all expressions shown in Fig.~\ref{fig:reparam}(b). 

\section{Conclusion}
\label{sec:conclusion}
We propose an innovative geometric-photometric joint alignment approach for facial mesh registration through the utilization of differentiable rendering techniques, demonstrating robust performance under various facial expressions with visibility changes. Unlike previous methods, our semantic annotation-free approach does not require marker point tracking or pre-aligned meshes for training. It is fully automatic and can be executed on a consumer GPU. Experiments show that our method achieve high geometric accuracy, surpassing conventional ICP-based techniques and the state-of-the-art method NPHM. We also validate the photometric alignment by comparing rendered images with captured multiview images, demonstrating pixel-level alignment in key facial areas, including the eyes, mouth, nostrils, and even freckles.

Our method currently adopts a relatively simple lighting and reflection model, which limits its performance under varying lighting conditions and skin highlights. Additionally, features such as teeth and the tongue may occasionally lead to inaccuracies in aligning the mouth contours. Future improvements will focus on optimizing efficiency, enhancing the rendering function to simulate more complex effects, and implementing cross-subject alignment based on genus-0 properties.



\bibliographystyle{elsarticle-num} 
\bibliography{manuscript_main}

\end{sloppypar}
\end{document}